\documentclass[12pt]{iopart}
\usepackage{psfig}
\usepackage{epsfig}
\usepackage{bm}
\newcommand{\be}{\begin{equation}}
\newcommand{\ee}{\end{equation}}

\newcommand{\re}[1]{Eq.~(\ref{#1})}

\newcommand{\ds}{\displaystyle}

\newcommand{\hsp}{\hspace*{1pt}}

\newcommand{\bea}{\begin{eqnarray}}
\newcommand{\eea}{\end{eqnarray}}

\begin{document}
\title[Summary of theoretical contributions]
{Summary of theoretical contributions}

\author{H.~St\"ocker$^{1,2}$}

\address{$^1$ Frankfurt Institute for Advanced Studies (FIAS),
 Johann Wolfgang Goethe University,
 Max-von-Laue-Str. 1,
 60438 Frankfurt am Main, Germany}
\address{$^2$ Institut f\"{u}r Theoretische Physik,
 Johann Wolfgang Goethe Universit\"{a}t,
 Max-von-Laue-Str. 1,
 60438 Frankfurt am Main, Germany}


\begin{abstract}
Results from various theoretical approaches and ideas
presented at this exciting meeting are reviewed. I also  point towards future
directions, in particular hydrodynamic behaviour induced by jets
traveling through the quark-gluon plasma, which
might be worth looking at in more detail.
\end{abstract}

\section{Theoretical overview}

We have witnessed an exciting conference with an excellent program,
heated scientific discussions and lots of new data and theoretical
ideas. Our topics ranged from astrophysics to field theory, from
heavy-ion reaction phenomenology to  big-bang cosmology.

My task, i.e. to review all in all about 30 theory talks, is combined
in this paper with a cross-disciplinary analysis of experiments, which
verify - pardon, falsify the theoretical conjectures in many cases -
there is an old saying that a theory can never be verified: even if
lots of data support the theory, at some point the theory will always
go astray...

Let me rearrange the order of the theory talks on the topics of our
meeting\footnote{Instead of giving explicit references I refer the
reader to the electronic proceedings available on the Web}:

\begin{itemize}
\item{Equation of State}
\item{Collective Dynamics}
\item{Jets : Production and Quenching}
\item{   Results from $p+p$, $p(d) +A$ and $A+A$ collisions}
\item{Signatures of Quark Gluon Plasma}
\item{QCD at Finite Temperature and Density}
\item{Multiparticle production, fluctuations and correlations}
\item{Cosmological Implications of the QCD Phase Transition}
\item{QCD Phenomenology}
\item{Low $x$ behaviour of QCD}
\item{ Strangeness and heavy flavor production}
\end{itemize}

The  common interest is given by the titel of the conference:
"{\it Physics and astrophysics of the quark gluon plasma}"\footnote{Somewhat in
this summary, also pardon the fact that some of the many
interesting items have not been taken up here, because I did not
witness the first days of the conference.}

\begin{itemize}
\item{Astrophysics}
\item{Lattice}
\item{Colored Glass}
\item{Fluctuations \& DCCs}
\item{J/Psi \& EM Probes}
\item{Strangeness}
\item{Transport Theory}
\item{Hydro \& Jets}
\end{itemize}

John Ellis  gave a beautiful survey of the common issues in both heavy-ion
physics and the big bang cosmology:  We do in both cases study a very
fast expansion of dense/hot strongly interacting matter, and do have
the task to reconcile whatever happened in the first few nanoseconds of
the big bang from the sparse debris found nowadays. The connection to
the matter-anti matter asymmetry problems is particularly exciting for
future topical studies at the LHC.  This is quite analogous to the
transient 6-8 fm/c $\approx  2-2.5 \times10^{-23} s$ timescale of the
collision processes at RHIC.

\begin{figure}[t]
\vspace*{5mm} \centerline{\psfig{file=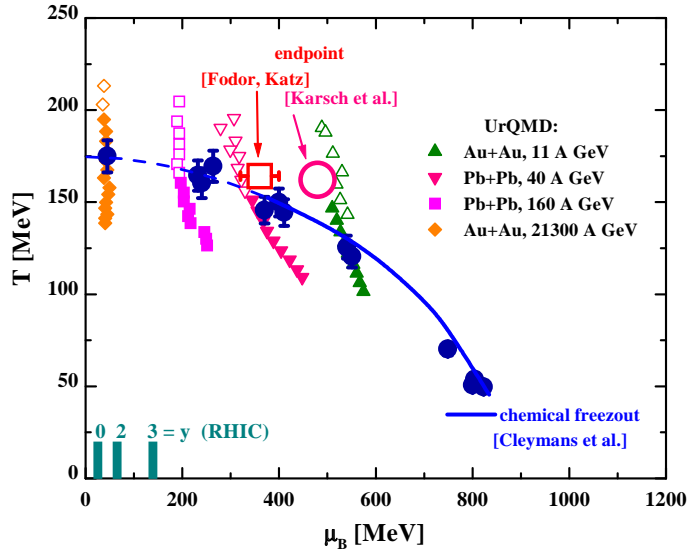,width=9cm}}
\caption{ The  phase diagram with the critical end point at $\mu_B
\approx 400 \mbox{ MeV}, T \approx 160 \mbox{ MeV} $ as predicted by
the Swansea-Bielefeld and Wuppertal/Budapest collaborations
\protect\cite{Fodor04}.  In addition, the time evolution in the
$T-\mu$-plane of a central cell in UrQMD calculations \cite{Bravina}
is depicted for different bombarding energies.  Note, that the
calculations indicate that bombarding energies $E_{LAB}
\stackrel{<}{\sim} 40$ A$\cdot$GeV are needed to probe a first order
phase transition as predicted by the Swansea-Bielefeld and
Wuppertal-Budapest collaborations. At RHIC (see insert at the
$\mu_B$ scale) this point is accessible in the fragmentation region
only (taken from \protect\cite{Bratkov04}). The new conjecture by
Gavai and friends is that the critical endpoint is moving to the
left to $T=0.95\  T_c$, $\mu_B = 1.1-1.3 T_c \approx$ 190--220 MeV.
In this case the top SPS energy range would be best suited to
explore the endpoint in central Au+Au collisions.} \label{phasedia}
\end{figure}

The intense astrophysics discussions between Bombaci and Banyopadyhyay
about the possible occurence of massive strange quark stars (SQS), the
transition of neutron stars to strange hyperon-, hybrid- and quark
stars, and the relation to the gamma ray bursts (quark-deconfinement
nova-model) has been of particular interest - this transition is
predicted to yield a radius-collapse of several kilometers.

The first observation of the "double delight pulsar" psr-j0737-3039
will enable  us to pin down the mass-radius curves by the spin-orbit
effect with high precision. D.  Bandyopadyhyay showed that soft
equations of state (EOS) are ruled out by EXO 0748-676.  The connection
of conjectured different color superconducting phases to the cooling
curves of SQS have been pointed out in the paper by Mishra and Mishra.

The lattice-QCD (lQCD) discussions between Gavai and Laermann centered
about the questions on the order of the phase transition and on the
speed of sound. Laermann stated that there is no indication for
criticality, while Gavai and friends showed that the critical endpoint
is at $T=0.95\ T_c$, $\mu_B/T = 1.1-1.3$, i.e.  less than half of the
$\mu_B$=400 MeV values given by the Swansea-Bielefeld and
Wuppertal-Budapest collaborations (cf. Fig 1). Gavai showed also that
all the way up to $T= 2 T_c$, the speed of sound is much less, $c_s^2=
0.15$ at $T = 1.1\ T_c$, than that of a noninteracting
ultrarelativistic (massless) gas, $c_s^2 = 1/3$.

\begin{figure}[t]
\centerline{\psfig{file=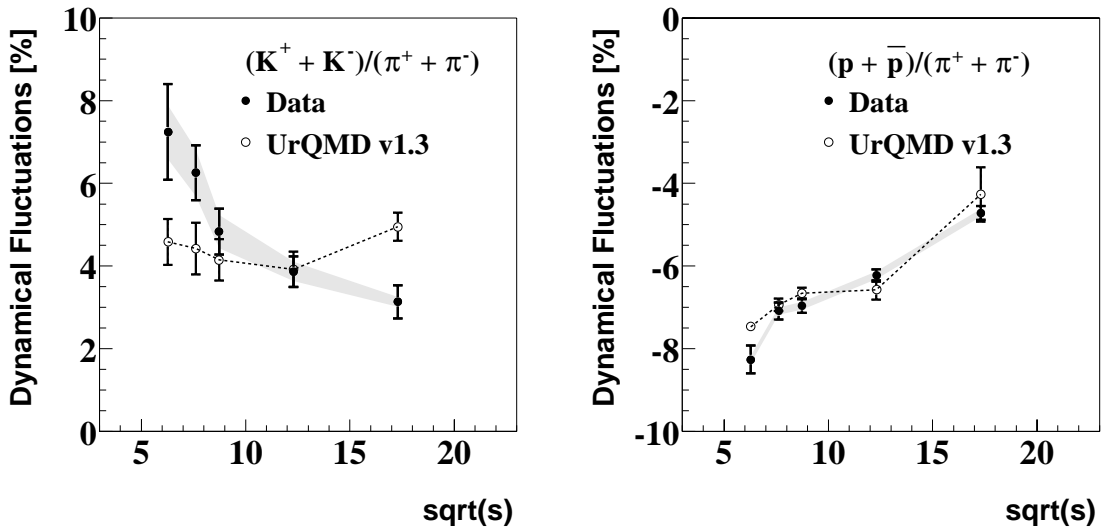,width=15cm}} \caption{ Energy
dependence of the event-by-event fluctuation signal of the
$(K^++K^-)/(\pi^++\pi^-)$ ratio (left panel) and the
$(p+\bar{p})/(\pi^++\pi^-)$ ratio (right panel).  The systematic
errors of the measurements are shown as grey bands (from Ref.
\cite{Roland}).} \label{fig2} \end{figure}

In the colored glass condensate section, Venugopalan explored the
demise of the structure function, in particular
how the dipole and higher multipole operators may turn out to be the
more relevant observables at high energies. Adding valence quark
contributions, Kovchegov showed a quite satisfactory agreement of the
Color-Glass-Condensate (CGC)-model to the observed rapidity dependence
of the $p_T$-distributions. McLerran iterated the theme of the Color
Glass Condensate as THE Medium: Pomerons, Odderons, Reggeons as
Quasiparticle excitations of the CGC - does this mean that the CGC is
the initial phase for the QGP?  Is the strong Quark-Gluon Plasma (sQGP)
really the CGC? Is rapid 'thermalization' due to the CGC? Does flow
arise largely from the CGC? Well, definitely  LHC is
THE CGC machine -- according to McLerran.

Fluctuations and Disordered Chiral Condensates (DCC's) were discussed
by Koch, Cs\"org\"o, Chandrasekar and Randrup, among others. $K/\pi$
fluctuations increase towards lower beam energy with a significant
enhancement over the hadronic cascade model UrQMD \cite{UrQMD} (cf.
Fig. 2)!  On the other hand, $p/\pi$ fluctuations are negative -- this
indicates a strong contribution from resonance decays, as was shown by
Koch in comparing NA49-data to UrQMD results.

Dileptons, J/Psi, and photons have been discussed by Lee, Mustafa and
Koch (among others). Large corrections on the QCD NLO  Quarkonium-
Gluon/hadron dissociation cross section have been reported even for the
Ypsilon system, especially near threshold. The thermal width of the
$J/\Psi$ should be $\sim$ 1 GeV at  T=600 MeV according to Lee's estimates.

\begin{figure}[t]
\centerline{\psfig{file=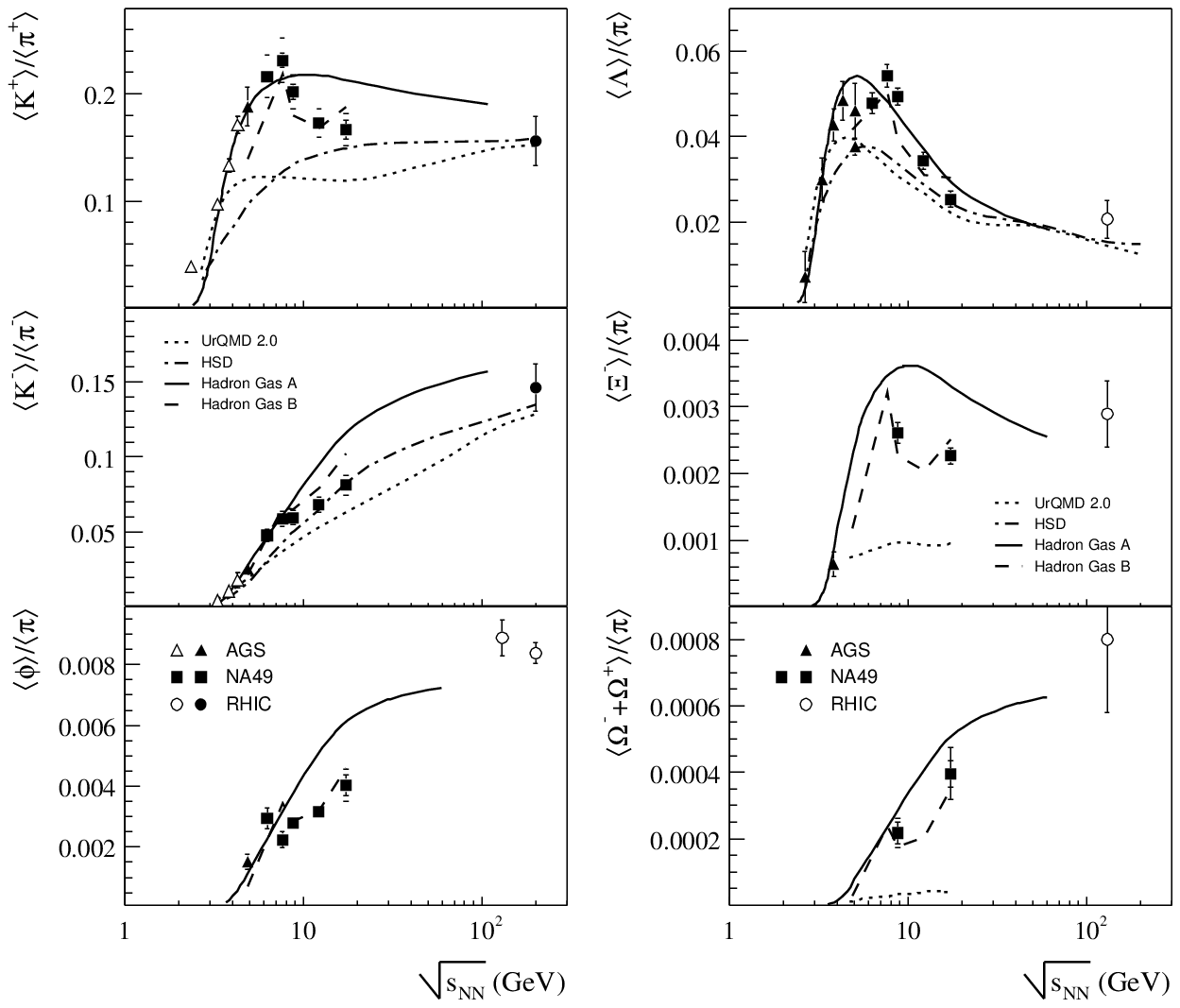,width=13.8cm}}
\vspace*{-3mm}
\caption{ The energy
dependence of the $4\pi$-yields of strange hadrons, normalized to the
pion yields, in central Pb+Pb/Au+Au collisions. The data are compared
to string hadronic models UrQMD2.0 \cite{UrQMD}: dotted lines;
HSD\cite{HSD}: dashed-dotted lines) and statistical hadron-gas models
from Braun-Munzinger and Becattini and collaborators (with strageness
undersaturation: dashed line, assuming full equilibrium: solid line).
The figure is taken from \cite{Blume}.} \label{fig3}

\vspace*{5mm}
\centerline{\psfig{file=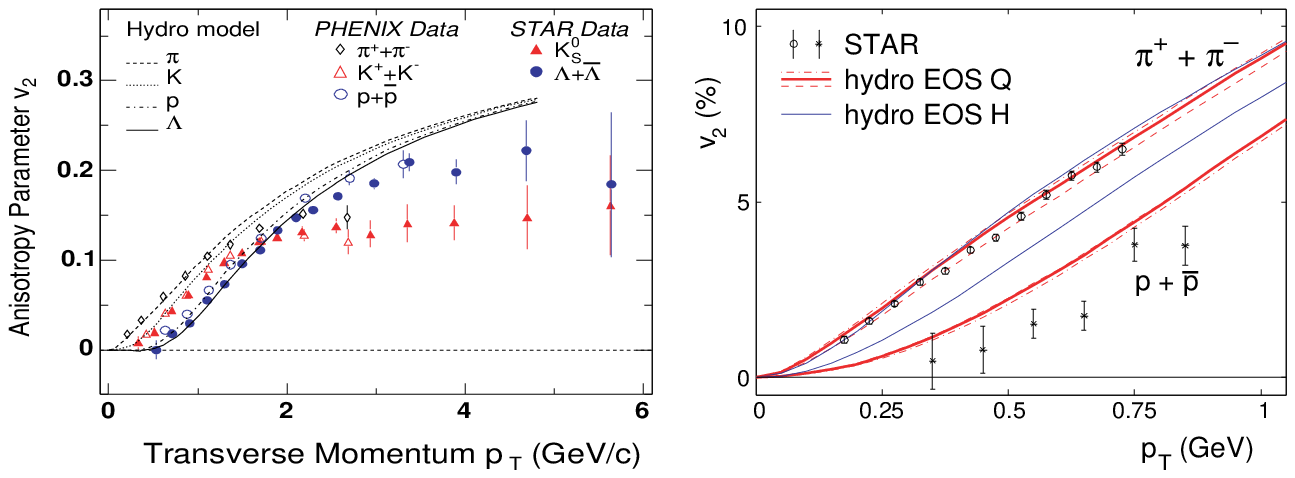,width=14cm}}
\caption{
Differential elliptic flow $v_2(p_\perp)$ for several identified hadron
species from minimum bias Au+Au collisions at $\sqrt{s}$= 130 GeV
(right) and $\sqrt{s}$ = 200 GeV compared with hydrodynamic predictions
from \protect\cite{Huovinen}. The figure is taken from \cite{heinz}.}
\label{fig4} \end{figure}

Strangegeness and equilibration has been the main topic of Rafelski,
Cleymans, Braun-Munzinger and Bleicher.  The structure in the
$K/\pi$ ratios reported by NA49 near $\sqrt(s)$ = 8 GeV is not
reproduced  by any model (cf. Fig. 3), but Peter Braun-Munzinger
notes: the natural smearing is 3 GeV near that energy - how can the
'horn' then be so steep? Hadron-string models work well globally, as
Bleicher reports, but these models do NOT give MULTI-STRANGE
BARYONs! Is the alternative a four parameter nonequilibrium
thermal model, with $T, \mu, \Gamma_\mu, \Gamma_s$, by Rafelski et
al.?

The extreme density/temperature dependence of the characteristic
equilibration time, $\tau_{eq.} \sim T^{-60}$, was pointed out by
Braun-Munzinger, which implies that all particles freeze out at about
the hadronization time. According to Braun-Munzinger this might be due
to Carsten Greiner's conjecture of Hagedorn states as intermediate
doorway states.

Deeply  bound $\bar{p}$ and $K^-$  states as  gateway to  cold  and
dense  matter were discussed by Walter Greiner: $\bar{p}$'s
 -- due to $G$-parity in the strong interactions -- and $K^-$ can
 suppress repulsive
vector fields, thus predicting discrete bound states with binding
energies of several 100th MeV and 20 fm/c life times \cite{Akaishi}.
Formation of such cold and highly dense nuclear system at densities
$\rho ~ 3-5 \rho_0$ will be studied in dense $\bar{p}$ -  nuclear
systems at FAIR (GSI)and the $K$-nucleus collisions at J-PARC.

Jacak, Shuryak, Heinz and Chauduri discussed applications of
hydrodynamics to RHIC-collisions. The reasons why hydro does reasonably
well fit both, radial and elliptic flow for a large number of hadron
species (cf. Fig. 4), is still not fully settled.  The question of
early thermalization and the unsatisfactory rapidity distributions from
ideal hydrodynamics remain open.

\begin{figure}[t]
\centerline{\psfig{file=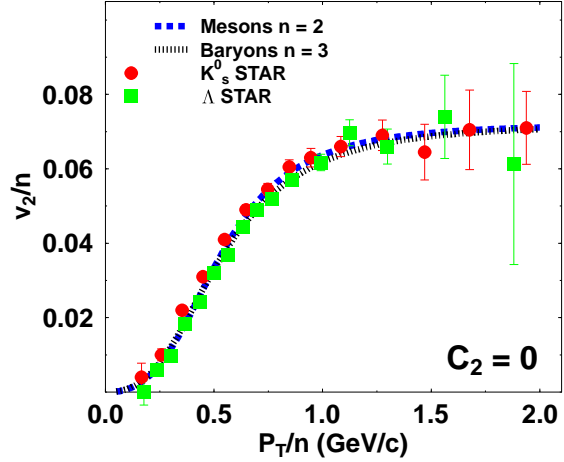,width=8cm}} \caption{ Scaled
elliptic flow $v_2/n(p_\perp/n)$ of baryons and mesons as calculated
from the quark flow compared to data for $\Lambda$-hyperons and
kaons. The figure is taken from \cite{Bass}.} \label{fig5}
\end{figure}

\begin{figure}[h]
\centerline{\epsfig{file=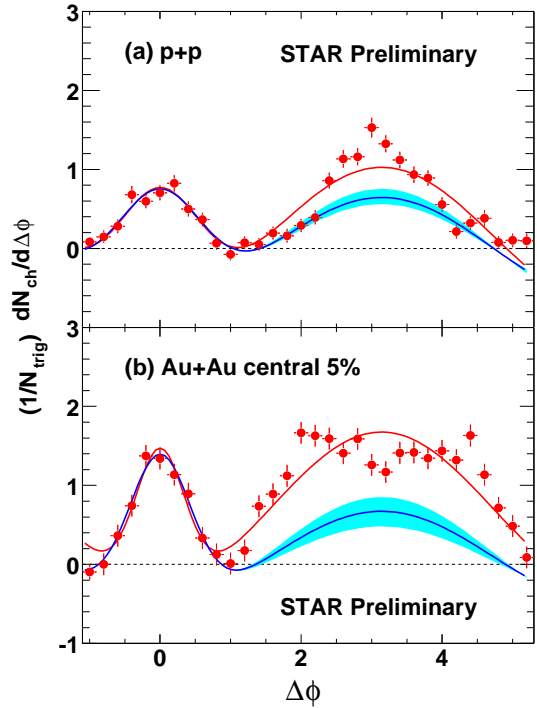,scale=0.4}}
\caption{ The per trigger particle normalized $\Delta \phi$
distributions for p+p (a) and 5\% most central Au+Au collisions (b).
The figure is taken from \protect\cite{STAR}.}
\label{fig6} \end{figure}

Bass showed in his talk, however, that the recombination/quark
coalescence models (cf. Fig. \ref{fig5}) can help analyze the
participant scaling and even the charm flow. However, as Bleicher
showed, even the hadron/string model UrQMD  may exhibit
"recombination" and participant scaling.

Jacak showed the PHENIX jet-pair distributions, which clearly give a
novel signal to the away-side jet suppression (cf. Fig. \ref{fig6} for
STAR results), i.e. the recent topic of Mach-cones induced by stopped
jets in the quark-gluon liquid \cite{Sto04}. This is most important as
an observable, because it links  the parton dynamics and collective
flow and the jet tomography to the measurement of the speed of sound in
the medium - be it a weakly or strongly coupled plasma: the opening
angle of the Mach-shock-wakes directly gives the speed of sound in the
medium, which is linked to both, the appearance of vector potentials
and the parton/constituent mass parameters.

\section{Interlude on Mach shocks}

Sideward peaks around the away-side jet have been predicted recently
\cite{Sto04} as a signature of Mach shock waves created by stopping
 partonic jets propagating through a QGP formed in an ultrarelativistic
heavy--ion collision. Analogous Mach shock waves were studied long ago
for heavy-ion induced Mach shocks travelling through cold hadronic
matter~\mbox{\cite{Hof74,Sto80}} as well as in nuclear Fermi
liquids~\cite{Gla59,Kho80}. It has been argued that Mach--like motions
 of quark--gluon matter can appear via the excitation of collective
plasmon waves by the moving color charge associated with the leading
jet \cite{Sto04,Schafer78}.

\begin{figure}[t]
\centerline{\includegraphics[width=0.5\textwidth]{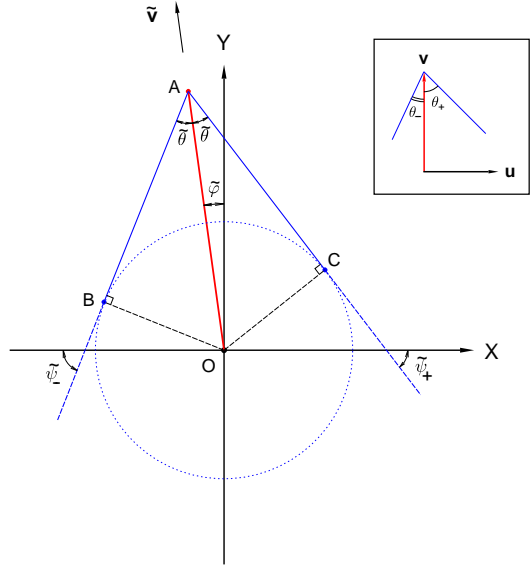}}
\caption{Mach region created by jet
moving with velocity $\bm{v}$ orthogonal to the fluid velocity~$\bm{u}$\,.
Main plot and insert correspond to FRF and CMF, respectively.
It is assumed that jet moves from $O$ to $A$ in FRF. Dotted
circle represents the front of sound wave generated at point $O$.}
\label{Mach2}
\end{figure}

Point--like perturbations (a small body, a hadron or parton etc.)
moving with a supersonic speed in the spatially homogeneous ideal fluid
produce the  Mach region of the perturbed matter \cite{Lan59}. In the
fluid rest frame (FRF) the Mach region has a conical shape (cf. Fig.
\ref{Mach2}) with an  opening angle with respect to the direction of
particle propagation given by\footnote{ Here and below the quantities
in the FRF are marked by a tilde.}
\begin{equation}
\label{mac1}
\tilde{\theta}_M=\sin^{-1}\left(\frac{c_s}{\widetilde{v}}\right) \, ,
\end{equation}
where $c_s$ denotes the sound velocity of the unperturbed (upstream)
fluid and $\tilde{\bm{v}}$ is the particle velocity with respect to the
fluid. In the FRF, trajectories of fluid elements (perpendicular to the
surface of the Mach cone) are inclined at the angle
$\Delta\theta=\pi/2-\widetilde{\theta}_M$  with respect to
$\widetilde{\bm{v}}$\,.
Strictly speaking, formula (\ref{mac1}) is applicable only for weak,
sound--like perturbations and certainly not valid for space--time
regions close to a leading particle.  Nevertheless, it suffices for a
qualitative analysis of flow effects. Following
Refs.~\cite{Sto04,Satarov,Cas04} one can estimate the angle of
preferential emission of secondaries associated with a fast jet in the
QGP. Substituting $\widetilde{v}=1, c_s=1/\sqrt{3}$ into~\re{mac1}
gives the value $\Delta\theta\simeq$ 0.96 rad = 61$^o$\.  This agrees
well with positions of maxima of the away--side two--particle
distributions observed by the STAR Collaboration (cf. \ref{fig6}) in
central Au+Au collisions at RHIC energies (cf. also B. Jacak's talk).

\begin{figure*}[t]
\centerline{\psfig{file=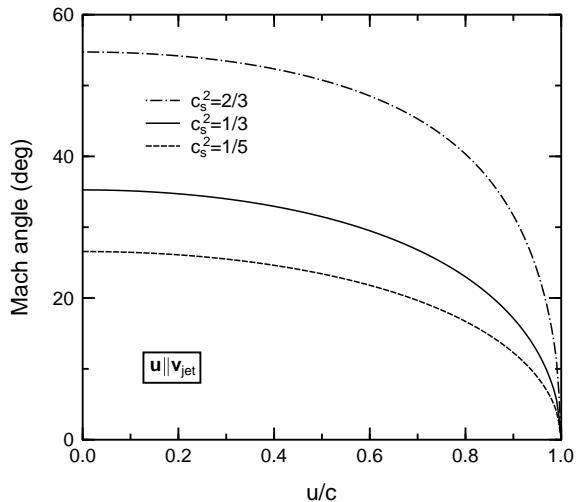,width=8cm}}
\caption{Mach cone angles for jet propagating collinearly to the
matter flow as a function of fluid velocity $u$\,. Different curves
correspond to different values of sound velocity $c_s$\,.}
\label{fig1}
\end{figure*}

Let us consider \cite{Satarov} the case when the away--side jet
propagates with velocity $\bm{v}$ parallel to the matter flow velocity
$\bm{u}$\,.  Assuming that $\bm{u}$ does not change with space and
time, and performing the Lorentz boost to the FRF, one sees that a weak
Mach shock has a conical shape with the axis along $\bm{v}$\,. In this
reference frame, the shock front angle $\widetilde{\theta}_M$ is given
by (\ref{mac1}).  Transformation from the FRF to the c.m. frame (CMF)
shows that the Mach region remains conical, but the Mach angle becomes
smaller in the CMF:
\be \label{macp1}
\tan{\theta_M}=\frac{\ds 1}{\ds\gamma_u}\tan{\widetilde{\theta}_M}\,,
\ee
where $\gamma_u\equiv (1-u^2)^{-1/2}$ is the Lorentz factor
corresponding to the flow velocity $\textbf{u}$\,.
The resulting expression for the Mach angle in the CMF is
\be \label{macp2}
\theta_M=\tan^{-1}
\left(c_s\sqrt{\frac{1-u^2}{\widetilde{v}^{\hsp 2}-c_s^2}}\right)\,,
\ee
where
\be \label{vrel}
\widetilde{v}=\frac{v\mp u}{1\mp v\hsp u}\,,
\ee
and upper (lower) sign corresponds to the jet's motion in (or opposite to)
the direction of collective flow. For ultrarelativistic jets ($v\to 1$) one
can take $\widetilde{v}\simeq 1$ which leads to a simpler expression
\be \label{macp3}
\theta_M\simeq\tan^{-1}\left(\frac{\ds c_s\gamma_s}{\ds \gamma_u}\right)=
\sin^{-1}\left(c_s\sqrt{\frac{1-u^2}{1-u^2\hsp c_s^2}}\right)\,,
\ee
where $\gamma_s=(1-c_s^2)^{-1/2}$\,. According to (\ref{macp3}),
in the ultrarelativistic limit $\theta_M$ does not depend on the
direction of flow with respect to the jet. The Mach cone
becomes more narrow as compared to jet propagation in static matter. This
narrowing effect has a purely relativistic origin. Indeed, the
difference between $\theta_M$ from (\ref{macp3}) and the Mach angle in
absence of flow ($\lim\limits_{u\to 0}{\theta_M}=\sin^{-1}{c_s}$) is of
 second order in the collective velocity $u$\,. The Mach angle
calculated from~(\ref{macp3}) is shown in Fig.~\ref{fig1} (from
\cite{Satarov}) as a function of $u$ for different sound velocities
$c_s$\,. Following Ref.~\cite{Cas04}, the value $c^2_s=1/5$ is
identified with the hadronic matter and $c^2_s=1/3$ with ideal QGP
composed of massless quarks and gluons. The value $c^2_s=2/3$ may be
chosen to represent a strongly coupled QGP~\cite{Shu04}.  We see that
precise measurements will provide valuable information on the
properties of the quark-gluon liquid \cite{Peshier-WC,Sto04}.

\begin{figure}[t]
\centerline{\psfig{file=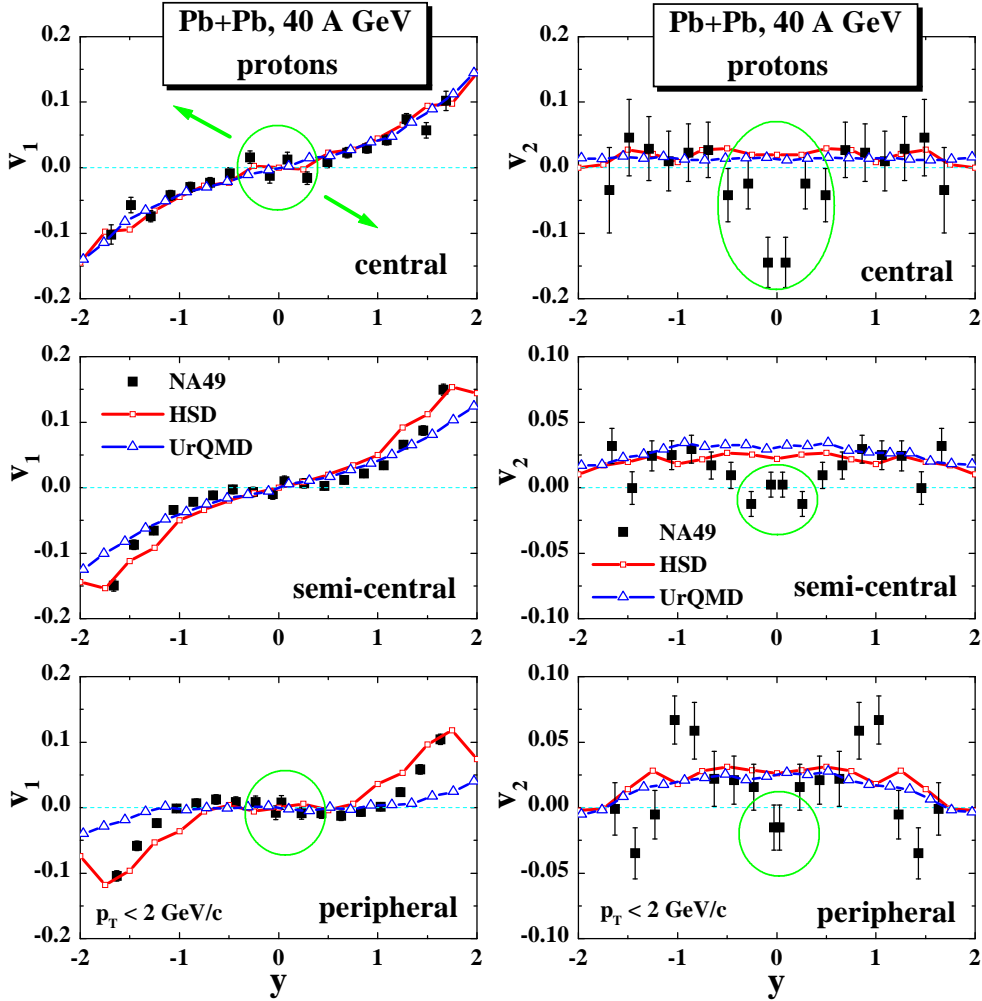,width=13cm}}
\caption{ The flow $v_1$ and $v_2$ for protons from NA49 \cite{NA49}
for Pb+Pb at 40 A GeV in comparison to the results of the hadron/string
models HSD (red lines) and UrQMD (blue lines). Note the large $v_2$
deviations (at $y_{cm}$) of the data from the best conventional
hadronic transport theories.  The figure is taken from
\protect\cite{LENA}.}
\label{fig7} \end{figure}

\section{Future directions}
I propose future correlation measurements which can yield
spectroscopic information on the plasma:
\begin{enumerate}

\item
Measure the sound velocity of the expanding plasma by the
emission pattern of the plasma particles traveling sideways with
respect to the jet axis: The dispersive wave generated by the wake of
the jet in the plasma yields preferential emission to an angle
(relative to the jet axis) which is given by the ratio of the leading
jet particles' velocity, devided by the sound velocity in the hot dense
plasma rest frame.  The speed of sound for a non-interacting gas of
relativistic massless plasma particles is $c_s \approx
\frac{c}{\sqrt{3}} \approx 57 \% \,c$, while for a plasma with strong
vector interactions, $c_s \approx c$, since strong shocks can yield larger speeds.
They are also related -- unlike the linearized sound waves --
to strong matter flow with high flow velocities $v_f$ approaching
the speed of light relative to the expanding medium.
Hence, the emission angle measurement
can yield information of the interactions in the plasma.

\item
The NA49 collaboration has observed the collapse of both, $v_1$- and
$v_2$-collective flow of protons (cf. Fig. \ref{fig7}), in Pb+Pb
collisions at 40 A$\cdot$GeV, which presents first evidence for a
first order phase transition in baryon-rich dense matter. It should
be possible to study the nature of this transition and the
properties of the expected chirally restored and deconfined phase
both at the forward fragmentation region at RHIC, with upgraded
and/or second generation detectors, and at the new GSI facility
FAIR.

\item
A critical discussion of the use of collective flow as a barometer
for the equation of state (EoS) of hot dense matter at RHIC showed
that hadronic rescattering models can explain $< 30 \%$ of the
observed elliptic flow $v_2$ for $p_T > 2$ GeV/c \cite{Gal05,Gal04}.
I interpret this as evidence for the production of superdense matter
at RHIC with initial pressure way above hadronic pressure, $p >
1$~GeV/fm$^3$.

\item
The fluctuations in the flow, $v_1$ and $v_2$, should be measured.
Ideal Hydrodynamics predicts that they are larger than 50 \%  due to
initial state fluctuations.  The QGP coefficient of viscosity may be
determined experimentally from the fluctuations observed and proof the
conjecture of Ref. \cite{Peshier-WC}.

\begin{figure}[t]
\begin{minipage}[r]{7 cm}
\psfig{file=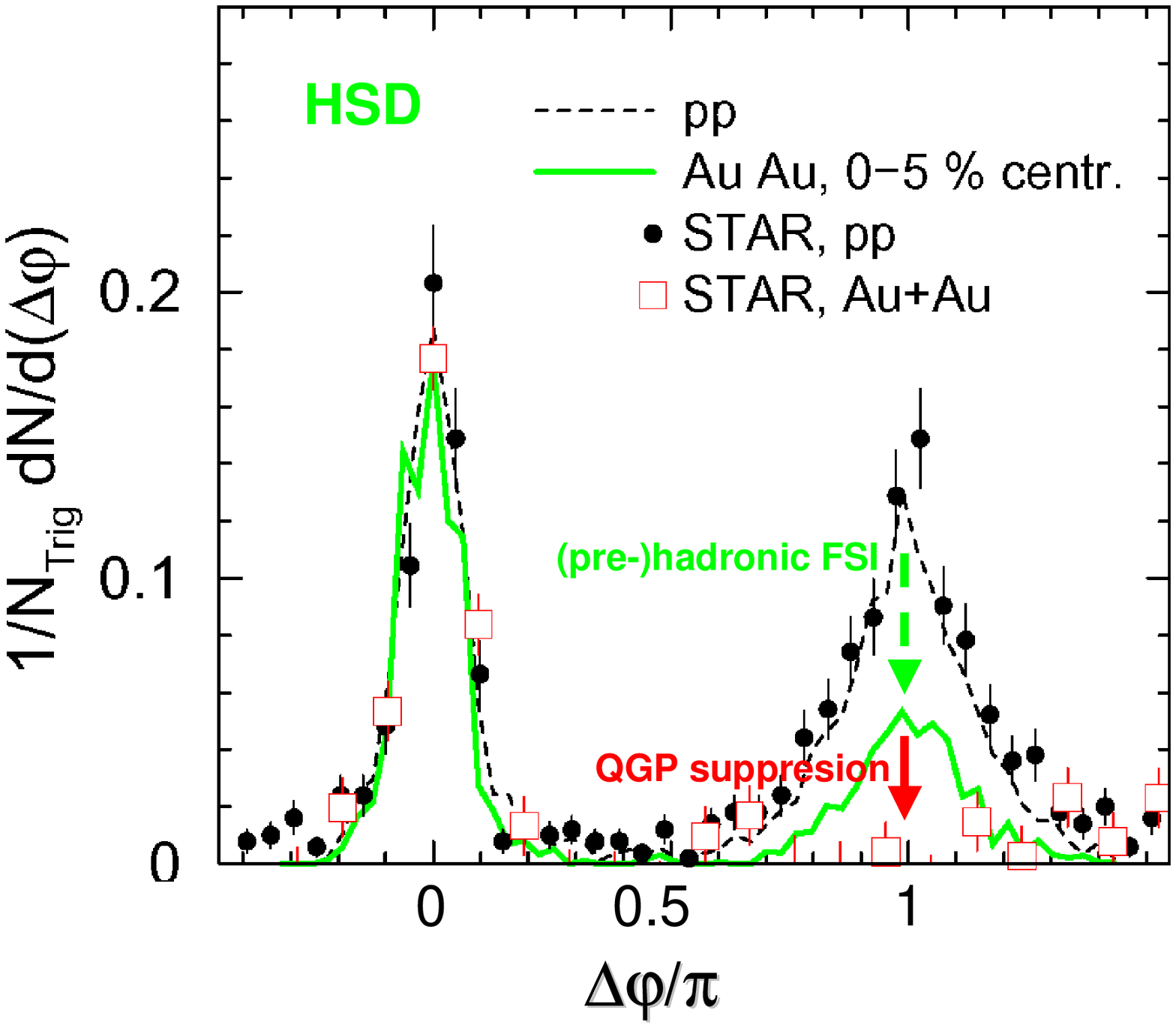,scale=0.38}
\end{minipage}
\begin{minipage}[r]{7 cm}
\vspace*{-2mm} \hspace*{1.2cm}
\epsfig{file=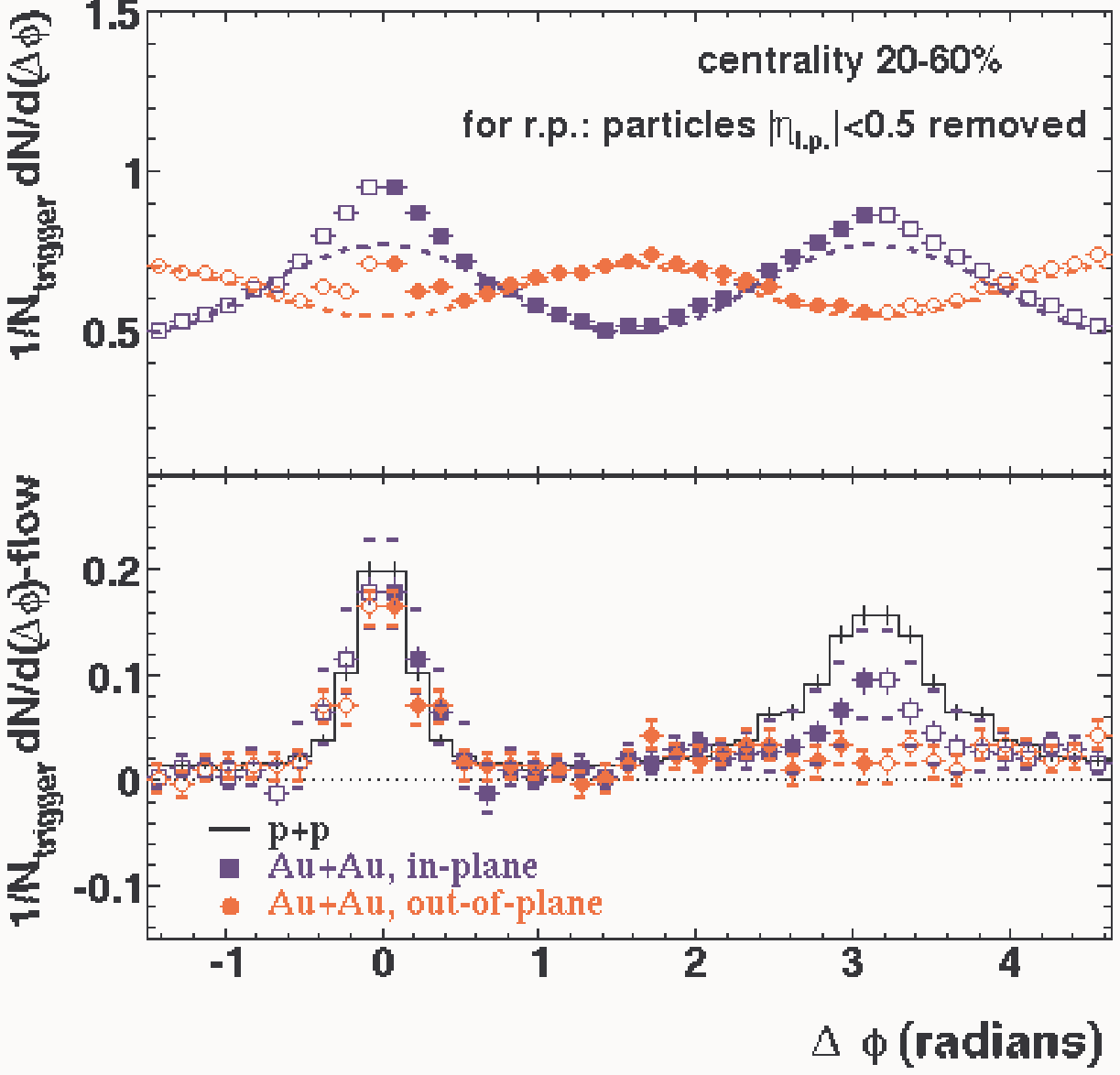,scale=0.5}
\end{minipage}
\caption{
Left: STAR data on near-side and away-side jet correlations compared
to the HSD model for p+p and central
Au+Au collisions at midrapidity for $p_T^{Trig}=4\dots6\,{\rm GeV}/c$ and
$p_T=2\,{\rm GeV}/c\dots p_T^{Trig}$ \protect{\cite{Gal04,Cassing04}}.
Right: High $p_T$ correlations: in-plane vs. out-of-plane
correlations of the probe (jet+secondary jet fragments) with the bulk
($v_2$ of the plasma at $p_T > 2\,$GeV/c), prove the existence of the
initial plasma state (STAR-collaboration, preliminary).}
\label{angcorr}
\end{figure}

\item
The connection of $v_2$ to jet suppression has  proven
experimentally that the collective flow is not faked by minijet
fragmentation and theoretically that the away-side jet suppression can
only partially ($<$ 50\%) be due to pre-hadronic or hadronic
rescattering \cite{Gal05} (cf. Fig. \ref{angcorr}).

\item
I propose upgrades and second generation experiments at RHIC, which
inspect the first order phase transition in the fragmentation region,
i.e. at $\mu_B\approx~200-400$~MeV  ($y \approx 3-5$), where the collapse
of the proton flow -- analogous to the 40 A$\cdot$GeV data -- should be seen.

\end{enumerate}

\vspace{0.5cm}
Let me finally express my birthday greetings to Bikash and thank him
and his crew for decades of exciting physics conjectures, his strong
involvement into our field and courage to built up such a great school
of young successful scientists in India, which are highly competitive
in the whole world.

\section*{Acknowledgments}
This work is partially supported by GSI, BMFT, DFG, and DAAD. Let
me thank Elena Bratkovskaya, Leonid Satarov and Igor Mishustin
(from FIAS) for their contributions to this Summary.

\end{document}